\documentclass[conference]{IEEEtran}

\usepackage{amsthm}
\usepackage[cmex10]{amsmath}
\usepackage{graphicx}
\usepackage{color}
\usepackage[tight,footnotesize]{subfigure}
\usepackage{amsfonts}
\usepackage{algorithmic}
\usepackage{algorithm}
\usepackage{graphicx}
\usepackage{subfigure}
\usepackage{bm}
\usepackage{times,color,amssymb,epsfig,psfrag,stfloats,enumerate}
\usepackage{cite}
\usepackage{flushend}

\newtheorem{definition}{Definition}

\begin{document}
%
\title{Energy-Age Tradeoff in Status Update Communication Systems with Retransmission}

\author{\IEEEauthorblockN{Jie Gong\IEEEauthorrefmark{1}, Xiang Chen\IEEEauthorrefmark{2} and Xiao Ma\IEEEauthorrefmark{1}\\}
\IEEEauthorblockA{\IEEEauthorrefmark{1} School of Data and Computer Science, Sun Yat-sen University, Guangzhou 510006, China}
\IEEEauthorblockA{\IEEEauthorrefmark{2} School of Electronics and Information Technology, Sun Yat-sen University, Guangzhou 510006, China}
Email: gongj26@mail.sysu.edu.cn}

\maketitle

\begin{abstract}
Age-of-information is a novel performance metric in communication systems to indicate the freshness of the latest received data, which has wide applications in monitoring and control scenarios. Another important performance metric in these applications is energy consumption, since monitors or sensors are usually energy constrained. In this paper, we study the energy-age tradeoff in a status update system where data transmission from a source to a receiver may encounter failure due to channel error. As the status sensing process consumes energy, when a transmission failure happens, the source may either retransmit the existing data to save energy for sensing, or sense and transmit a new update to minimize age-of-information. A threshold-based retransmission policy is considered where each update is allowed to be transmitted no more than $M$ times. Closed-form average age-of-information and energy consumption is derived and expressed as a function of channel failure probability and maximum number of retransmissions $M$. Numerical simulations validate our analytical results, and illustrate the tradeoff between average age-of-information and energy consumption.
\end{abstract}


%
\IEEEpeerreviewmaketitle

\section{Introduction}
In real-time monitoring and control applications for Internet of Things (IoT) such as temperature status sensing, power grid phase data update, velocity and acceleration monitoring of a vehicle, etc., timely information update is strictly required to guarantee the availability of the latest system state. Therefore, instead of transmission or queuing delay, people are interested in a novel performance metric termed as \emph{age-of-information} \cite{kaul2012real}, or simply \emph{age}, defined as the time elapsed since the generation of the latest received update. On the other hand, for status sensing applications, sensor nodes are usually battery powered, i.e., energy-constrained. It is critical to reduce energy consumption when updating status information timely. Hence, similar to energy-delay tradeoff in conventional wireless communications, there exists a tradeoff between energy consumption and age-of-information.

In the literature, the concept of age-of-information has been extensively studied. The average age for a first-come-first-served (FCFS) queuing system was analyzed in \cite{kaul2012real}, where information updates are generated randomly. When a source has access to the channel's idle/busy state, it can generate information updates by its own will. In this case, generating a fresh update just as the prior update is delivered and the channel becomes idle, known as \emph{just-in-time scheduling} \cite{yates2015lazy}, is preferred as it completely eliminates the waiting time in the queue. The optimality of just-in-time scheduling, or \emph{zero-wait policy}, was analyzed in \cite{sun2017update} with generalized age penalty functions. Concerning energy consumption, information age in energy harvesting systems was analyzed. The age-optimal policies under infinite battery, unit battery, and generalized finite battery were studied in \cite{bac2015age}, \cite{bac2017scheduling}, and \cite{bac2018achieving}, respectively, assuming that each packet can be delivered instantly using a unit energy. When the packet delivery time is taken into account and is inversely related to the transmit power, total age-of-information minimization problem subject to energy causality constraints was considered in \cite{arafa2017age}. However, the above works only consider transmission energy consumption, while ignore the energy for sensing which is not negligible in many applications \cite{liu2016energy}.

When considering sensing energy consumption, there is a tradeoff between energy and age-of-information when transmitting updates via an error-prone wireless channel. If a receiver fails to receive a status update packet and sends back a NACK to its transmitter, the transmitter needs to determine whether to retransmit the existing packet or to sense a fresh status and transmit a new packet. Without sensing energy consumption, obviously sensing and transmitting a new packet is the best choice \cite{elif2017average}. However, if sensing process consumes energy, retransmitting the existing packet can save energy consumption for sensing, while sensing and transmitting the new packet is expected to reduce the average age. The retransmission policies for uncontrolled status sensing process were studied in \cite{chen2016age, najm2017status}. However, to the best of our knowledge, there is no existing work dedicated to the analysis of energy-age tradeoff. Although both sensing energy and retransmissions were jointly considered in \cite{liu2016energy} and age performance was analyzed, the optimal energy-age tradeoff was still absent.

The energy-age tradeoff analysis can be originated to the studies on energy-delay tradeoff. The monotonic relation between transmit power and average delay in fading channels was analyzed in \cite{berry2002commun}. When jointly considering non-ideal circuit power, the tradeoff between energy and delay may not be always monotonic. To achieve the optimal tradeoff, sleeping policy should be adopted to match energy consumption with traffic requirement \cite{wu2013traffic, wu2016base}. In \cite{niu2015char}, a dedicated base station sleeping energy consumption model was derived, and energy-delay performance evaluation for $N$-policy was given. Despite of a large amount of studies on energy-delay tradeoff, age-of-information is a quite different performance metric compared with delay in terms of concepts and ways of calculation. Hence, energy-age tradeoff analysis is an open issue. Moreover, it is also important in status monitoring applications since sensors are usually energy-constrained.

In this paper, we study a status update system where a source detects environmental status and sends status updates to a receiver via an error-prone channel. Both sensing process and transmission process consumes energy. The receiver feeds back an ACK/NACK depending on whether reception is successful or not. Upon receiving a NACK, the source needs to decide either to detect and transmit a fresh status update or to retransmit the existing packet. We analyze the average age-of-information and average energy consumption of a threshold-based retransmission policy, i.e., the source retransmits the same packet no more than $M$ times. Numerical results illustrate the tradeoff between energy and age, as well as the influence of power control on the energy-age tradeoff.

\section{System Model}
\begin{figure}
\centering
\includegraphics[width=3.4in]{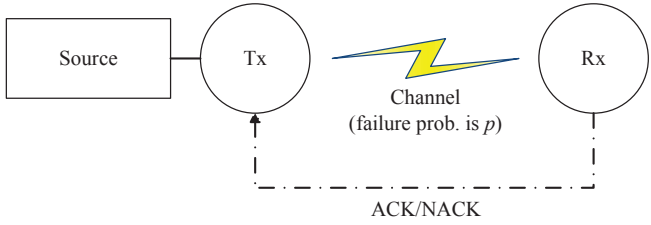}
\caption{Status update system model with transmission failure and ACK/NACK feedback for retransmission.} \label{fig:system}
\end{figure}

Consider a status update system as shown in Fig.~\ref{fig:system}. The latest status information is sensed and an update packet containing this information is generated by the source. Then, the update is transmitted from the transmitter to the receiver through an identically and independently distributed (i.i.d.) channel with transmission failure probability $p \in (0, 1)$. After a successful reception, the receiver sends an ACK signal back to the transmitter through a error-free feedback channel. Otherwise, the receiver sends back a NACK signal in case of a failure. If an ACK is sent back to the transmitter, it will notify the source to generate a new packet to update the status information. The energy consumption for status sensing is denoted by $E_s$. If a NACK is sent on the other hand, the current packet may be retransmitted to save energy for status sensing. Each packet is allowed to be transmitted no more than $M$ times. In our system, a slot is defined as the time duration from transmitting a packet to receiving an acknowledgement feedback. The slot length is denoted by $T$, and the energy consumption for transmitting a packet is denoted by $E_t$. Without loss of generality, we set $T = 1$ in the rest of this paper.

Age-of-information represents the freshness of data. At any time $t$, if the latest data packet that is successfully received by the receiver is generated at time $U(t)$, the age can be expressed as
\begin{align}
\Delta(t) = t - U(t).
\end{align}
In other words, the age-of-information is the time elapsed since the moment the freshest received update was generated.

\begin{figure}
\centering
\includegraphics[width=3.4in]{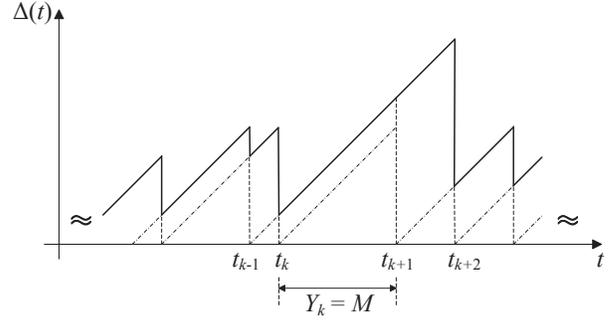}
\caption{Evolution of age-of-information.} \label{fig:aoi}
\end{figure}

The evolution of $\Delta(t)$ in this model is depicted in Fig.~\ref{fig:aoi}, where $t_k$ denotes the time instance when the $k$-th packet is generated. Denote $Y_k = t_{k+1} - t_k$ as the time interval between two consecutive packet generations. Recalling that the transmission slot length is normalized as $T=1$, $Y_k$ equals to the total number of transmissions of the $k\textrm{-th}$ packet. If the current transmission succeeds, the instantaneous age decreases and a new update will be generated. If the current transmission fails when $Y_k = M$, a new packet will be generated and transmitted while the age-of-information will continuously increase until a successful reception. The average age-of-information is typically defined as
\begin{align}
\bar{\Delta} = \lim_{\tau \rightarrow \infty} \frac{1}{\tau} \int_{0}^{\tau} \Delta(t) \mathrm{d} t.
\end{align}
In our model, it can be calculated over transmission slots and will be detailed in the next section. The average energy consumption can be calculated as
\begin{align}
\bar{E} = \lim_{k \rightarrow \infty} \frac{kE_s + \sum_{i=1}^{k} Y_iE_t}{\sum_{i=1}^k Y_i}. \label{eq:energy}
\end{align}
In the above equation, the denominator is the total number of transmission slots, and the nominator is composed of sensing energy consumption $kE_s$ and transmission energy consumption $\sum_{i=1}^{k} Y_iE_t$, where $E_t$ is the the energy consumption for transmitting one packet. It is modeled as a function of the transmit power $P_t$, i.e.,
\begin{align}
E_t = P_c + \eta P_t,
\end{align}
where $P_c$ is the circuit energy consumption, $P_t$ is the transmit power, and $\eta$ is the inverse of the drain efficiency of power amplifier. When tuning the transmit power $P_t$, the channel transmission failure probability changes accordingly. Assume the channel follows Rayleigh fading, and hence, the failure probability can be calculated as \cite[Eq.~5.55]{tse2005fund}
\begin{align}
p = 1 - \exp \left( -\frac{(2^R-1)\sigma^2}{P_t} \right), \label{eq:p}
\end{align}
where $R$ is the data rate, $\sigma^2$ is the noise power.

Due to the existence of transmission failures and retransmissions, the calculation of the average age-of-information is not trivial, which is detailed in the next section.

\section{Average age-of-information and Energy Consumption}
In this section, the main results are listed at first for ease of reading. Then the analysis on the average age-of-information and the average energy consumption is presented in detail.

\subsection{Main Results}
With constant transmit energy consumption $E_t$, the average age-of-information and the average energy consumption can be respectively expressed as
\begin{align}
\bar \Delta &= \frac{3+p}{2(1-p)} - \frac{Mp^M}{1-p^M}, \label{eq:delta}\\
\bar E &= \frac{1-p}{1-p^M}E_s + E_t, \label{eq:Ebar}
\end{align}
where $p$ is expressed as \eqref{eq:p}. It can be seen that with the increase of $M$, the average energy consumption decreases due to the reduced number of status sensing, while on the other hand, the average age-of-information may increase. Hence, there is a tradeoff between reducing total energy consumption and maintaining a low information age. The tradeoff can also be obtained by adjusting the transmit power to balance the transmit energy consumption and channel failure probability. It will be illustrated later in numerical result section.

\subsection{Calculation of Average Age-of-Information}
For ease of calculation, the system is re-indexed by the successfully received packets instead of the generated packets. In the rest of this paper, the index $k$ refers to the $k$-th successfully received packet which may contain multiple packet generations. For clarity, we present the following definitions.
\begin{definition}
Define $\tilde t_{k}$ as the time instance for the generation of a new update after the $k$-th successful reception.
\end{definition}
\begin{definition}
Define $\tilde Y_{k} = \tilde t_{k+1} - \tilde t_{k}$ as the time duration between two consecutive successful receptions.
\end{definition}
\begin{definition}
Define $\hat Y_{k}$ as the transmission time duration for the $k$-th successfully received packet.
\end{definition}

\emph{Remark}: Recall that $Y_k$ is the time duration between two consecutive packet generations. As some packets may not be successfully received due to transmission failure, $\tilde Y_{k}$ may contain several $Y_k$s. For instance, in Fig.~\ref{fig:aoi}, there are two packet generations at $t_k$ and $t_{k+1}$, respectively. The former is not successfully received while the latter is successfully received at $t_{k+2}$. Thus, we have $\tilde Y_{k'} = Y_k + Y_{k+1}$ for some $k'$. On the other hand, $\hat Y_{k}$ only concerns the successfully received packet $k$. In Fig.~\ref{fig:aoi}, we have $\hat Y_{k'} = Y_{k-1}$ for some $k'$. In the following, we use $\tilde Y_{k}$ and $\hat Y_{k}$ to derive the main results.

\begin{figure}
\centering
\includegraphics[width=3.4in]{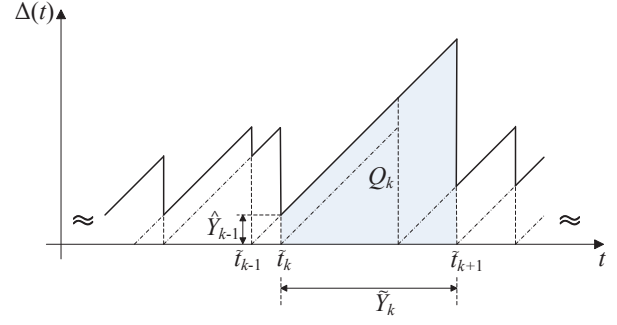}
\caption{Re-indexed evolution of age-of-information.} \label{fig:aoinew}
\end{figure}

By using the above definitions, the evolution curve of $\Delta(t)$ can be re-depicted as Fig.~\ref{fig:aoinew}. Then, the average age-of-information can be calculated as the average area of the gray trapezoid $Q_k$, i.e.,
\begin{align}
\bar{\Delta} = \lim_{k \rightarrow \infty} \frac{\sum_{i=1}^k Q_i}{\sum_{i=1}^k \tilde Y_i} = \frac{\mathbb{E} [Q_k]}{\mathbb{E} [\tilde Y_k]}. \label{eq:avgAoI2}
\end{align}

According to \eqref{eq:avgAoI2}, to calculate the average age-of-information, we need to calculate $\mathbb{E} Q_k$ and $\mathbb{E} \tilde Y_k$. Firstly, we have
\begin{align}
\mathbb{E} [Q_k] &= \mathbb{E} \left[ \frac{1}{2} (\hat Y_{k-1} + \tilde Y_k + \hat Y_{k-1}) \tilde Y_k \right] \nonumber\\
& = \mathbb{E} \left[ \hat Y_{k-1} \tilde Y_k \right] + \frac{1}{2}\mathbb{E} [\tilde Y_k^2] \nonumber\\
& = \mathbb{E} [\hat Y_{k-1}] \mathbb{E}[\tilde Y_{k}]  + \frac{1}{2}\mathbb{E} [\tilde Y_{k}^2], \label{eq:Qk}
\end{align}
where the first equality holds by the definition of a trapezoid's area, an the third equality holds as the transmission of each packet is independent. Combining \eqref{eq:avgAoI2} and \eqref{eq:Qk}, we have
\begin{align}
\bar{\Delta} = \mathbb{E} [\hat Y_k] + \frac{\mathbb{E} [\tilde Y_k^2]}{2\mathbb{E} [\tilde Y_k]}. \label{eq:avgAoI3}
\end{align}
Thus, to calculate $\bar{\Delta}$, we only need to focus on the distributions of $\tilde Y_k$ and $\hat Y_k$.

As the slot length is set to 1, the random variable $\tilde Y_k$ equals to the number of transmissions between two consecutive successful receptions. It follows a geometric distribution, i.e.,
\begin{align}
\mathrm{Pr}(\tilde Y_k = m) = p^{m-1}(1-p),
\end{align}
where $m = 1, 2, \cdots$. Then, we have
\begin{align}
\mathbb{E} [\tilde Y_k] &= \frac{1}{1-p}, \label{eq:Etildeyk}\\
\mathbb{E} [\tilde Y_k^2] &= \mathbb{E} [(\tilde Y_k - \mathbb{E} \tilde Y_k)^2] + (\mathbb{E} [\tilde Y_k])^2 \nonumber\\
&= \frac{p}{(1-p)^2} + \frac{1}{(1-p)^2} = \frac{1+p}{(1-p)^2}. \label{eq:Etildeyk2}
\end{align}

The random variable $\hat Y_k$ equivalents to the number of transmissions of the $k$-th successfully received packet. Hence, it has finite possible values, i.e., $\hat Y_k \in \{1, 2, \cdots, M\}$. Its distribution can be calculated based on the distribution of $\tilde Y_k$. In particular, $\hat Y_k = m$ corresponds to a set of events $\tilde Y_k = lM+m, \forall l$, which means consecutive $l$ failures and packet transmission plus $m$ retransmission before success. Therefore, we have
\begin{align}
\mathrm{Pr}(\hat Y_k = m) &= \sum_{l \ge 0} \mathrm{Pr} (\tilde Y_k = lM+m) \nonumber\\
&= \sum_{l\ge 0} p^{lM+m-1}(1-p) \nonumber\\
&= \frac{1-p}{1-p^M} p^{m-1},
\end{align}
where $m = 1, 2, \cdots, M$. Then by definition, we have
\begin{align}
\mathbb{E} [\hat Y_k] &= \sum_{m=1}^M m \mathrm{Pr}(\hat Y_k = m) \nonumber\\
& = \frac{1-p}{1-p^M} \sum_{m=1}^M m  p^{m-1} \nonumber\\
& = \frac{1}{1-p} - \frac{Mp^M}{1-p^M}. \label{eq:Ehatyk}
\end{align}

Combining \eqref{eq:avgAoI3}, \eqref{eq:Etildeyk}, \eqref{eq:Etildeyk2}, and \eqref{eq:Ehatyk}, the average age-of-information \eqref{eq:delta} can be obtained.

\subsection{Calculation of Average Energy Consumption}
The average energy consumption \eqref{eq:energy} can be rewritten as
\begin{align}
\bar E &= \lim_{k \rightarrow \infty} \frac{kE_s}{\sum_{i=1}^k Y_i} + E_t \nonumber\\
&= \lim_{k \rightarrow \infty} \frac{\sum_{i=1}^k g(\tilde Y_i)}{\sum_{i=1}^k \tilde Y_i}E_s + E_t \nonumber\\
&= \frac{\mathbb{E} [g(\tilde Y_k)]}{\mathbb{E} [\tilde Y_k]} E_s + E_t, \label{eq:energy2}
\end{align}
where $g(\tilde Y_k)$ is the number of status sensing in the interval $\tilde Y_k$ (including the status sensing at the beginning of the interval). In the case that the $l$-th packet is successfully received while the previous $l-1$ consecutive packets are not, $g(\tilde Y_k) = l$. It corresponds to $\tilde Y_k = (l-1)M+m, \forall m = 1, \cdots, M$. Hence, the distribution of $g(\tilde Y_k)$ is
\begin{align}
\mathrm{Pr}(g(\tilde Y_k) = l) &= \sum_{m=1}^M \mathrm{Pr} (\tilde Y_k = (l-1)M+m) \nonumber\\
&= \sum_{m=1}^M p^{(l-1)M+m-1}(1-p) \nonumber\\
&= p^{(l-1)M}(1-p^M),
\end{align}
where $l = 1, 2, \cdots$. Thus, the expectation can be expressed as
\begin{align}
\mathbb{E} [g(\tilde Y_k)] &= \frac{1}{1-p^M}. \label{eq:gyk}
\end{align}
Combining \eqref{eq:Etildeyk}, \eqref{eq:energy2}, and \eqref{eq:gyk}, the average energy consumption \eqref{eq:Ebar} can be obtained.

\section{Numerical Results}
In this section, numerical simulations are conducted to show the energy-age tradeoff performance. We adopt the power consumption model for home applications from EARTH project~\cite{imran2011energy}. In particular, $P_c = 2.1$ W, $\eta = 19.2308$, and $P_t \le P_{\max} = 20$ dBm. We evaluate the impact of $M$, $P_t$ as well as $E_s$ on the energy-age tradeoff curve.

\subsection{Fixed Transmission Energy}
Firstly, consider the case that $E_t = P_c + \eta P_{\max}$ is constant. We set $E_s = E_t$ and calculate the theoretical results for different settings of $p$ and $M$ according to \eqref{eq:delta} and \eqref{eq:Ebar}. In this simulation, the value of $p$ indicates the channel quality. To validate the theoretical results, we generate a large number of data packets and random channel realizations to simulate the status sensing and packet retransmission process, and then calculate average energy consumption and age-of-information.

\begin{figure}
\centering
\includegraphics[width=3.4in]{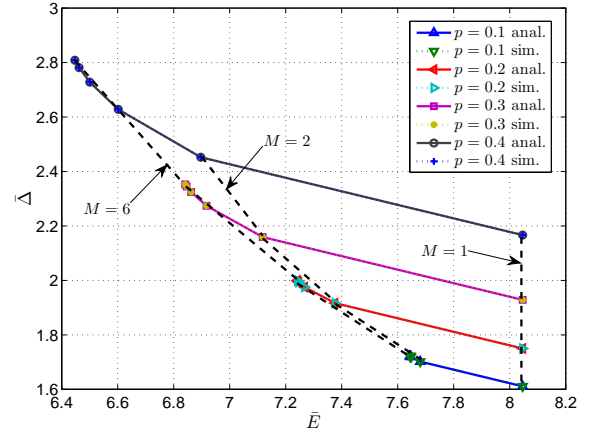}
\caption{Energy-Age tradeoff with constant transmit power. $E_s = E_t$.} \label{fig:thyVSsim}
\end{figure}

As shown in Fig.~\ref{fig:thyVSsim}, the numerical simulations match well with the analytical results, which validates our analysis. More importantly, for a given channel failure probability $p$, there is a tradeoff between energy consumption and age-of-information by varying the maximum number of allowed retransmissions $M$. For instance, with $p = 0.4$, the average energy consumption can be reduced by 1.6 via increasing $M$ from 1 to 6 while the average age-of-information is increased by 0.64. It is noticeable that with small value of $p$, the increase of $M$ does not have significant impact on the performance. In particular, with $p=0.1$, the results with $M = 3, \cdots, 6$ overlaps. This means that when the channel condition is good enough, the reception will succeed with high probability within a small number of transmissions. In addition, the maximum average energy consumption for all curves is 8.05, which corresponds to generating and then transmitting a new packet in each slot.

\begin{figure}
\centering
\includegraphics[width=3.4in]{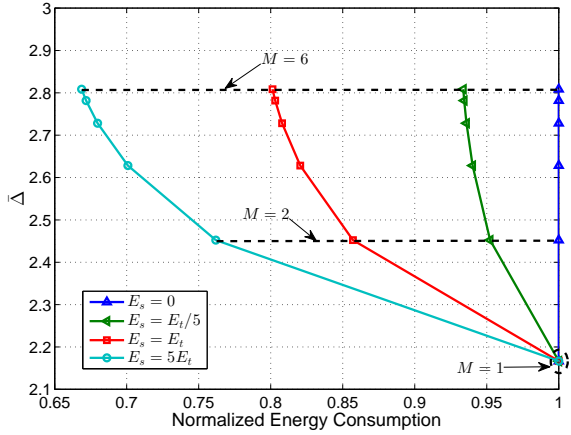}
\caption{Impact of sensing energy consumption on energy-age tradeoff with constant transmit power. $p = 0.4$.} \label{fig:AgeVSEs}
\end{figure}

Then, we change the value of $E_s$ to show its impact on the tradeoff. The result with $p = 0.4$ is shown in Fig.~\ref{fig:AgeVSEs}, where the energy consumption is normalized by $E_s + E_t$ for ease of comparison. It can be seen that when $E_s = 0$, the total energy consumption does not change. Hence, it is better to sense and transmit a new packet in each slot. While when $E_s$ is non-zero, the energy consumption can be reduced by trading the age performance. With the same average age, less energy is consumed if a larger $E_s$ is given. Therefore, retransmission is preferred when the sensing energy consumption is large.

\subsection{Transmit Power Control}
\begin{figure}
\centering
\includegraphics[width=3.4in]{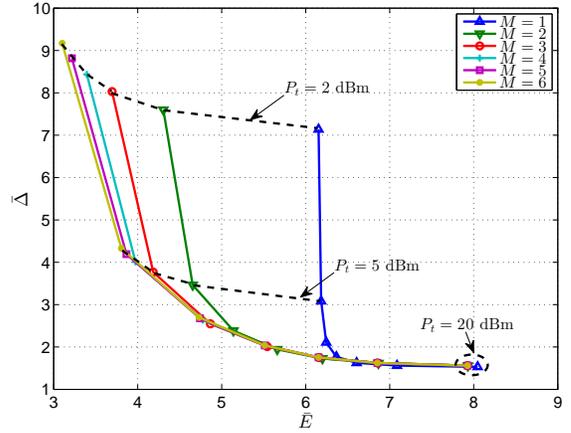}
\caption{Energy-Age tradeoff with transmit power control. $E_s = P_c + \eta P_{\max}$.} \label{fig:PCcurve}
\end{figure}

Next, we study the impact of transmit power control on the energy-age curve. The channel failure probability is determined according to \eqref{eq:p}, where the reference SNR with $P_{\max} = 20$ dBm is set to 20 dB to calculate $\sigma^2$, and the data rate $R = 2$ bps/Hz. We test the transmit power from 2 dBm to 20 dBm with sampling interval 3 dB. The result with fixed sensing energy $E_s = P_c + \eta P_{\max}$ is depicted in Fig.~\ref{fig:PCcurve}. It can be seen that for a fixed $M$, the average age-of-information decreases with the increase of the transmit power, while the energy consumption increases as well. For a fixed transmit power $P_t$, there is also a tradeoff between age-of-information and energy consumption. Even for $M=1$, i.e., a new packet is generated and transmitted in each slot, we still have a tradeoff between energy and age, despite that the curve is steep.

\begin{figure}
\centering
\includegraphics[width=3.4in]{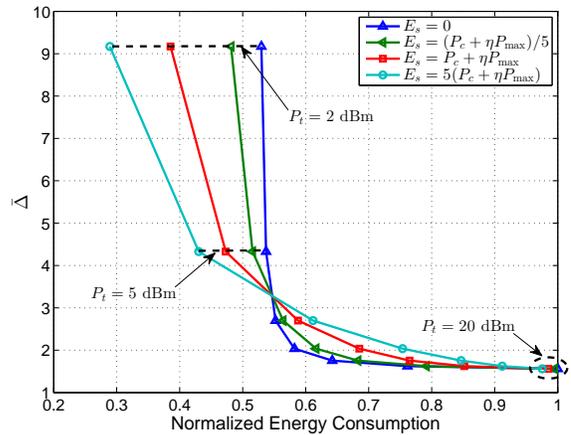}
\caption{Impact of sensing energy consumption on energy-age tradeoff with transmit power control. $M = 6$.} \label{fig:AgeVSEsPC}
\end{figure}

Finally, the impact of sensing energy on energy-age tradeoff with transmit power control is illustrated in Fig.~\ref{fig:AgeVSEsPC}, where we set $M=6$, and normalize the energy consumption by $E_s + (P_c + \eta P_{\max})$. Similar to the constant transmit power case, the tradeoff range is wider when the sensing energy is larger. The difference is that there still exists energy-age tradeoff when $E_s = 0$.

\section{Conclusion and Future Work}
In this paper, the tradeoff between age-of-information and energy consumption in the presence of channel error and retransmissions was studied. By deriving the closed form expressions and running the numerical simulations, the impact of different parameters on the energy-age tradeoff was extensively analyzed. The main conclusions include: (1) With the existence of sensing energy, there is a tradeoff between energy consumption and age-of-information by tuning the retransmission threshold $M$ to balance energy consumption for sensing and transmission. (2) The impact of retransmission policy is more significant when the channel failure probability is larger. (3) In the absence of sensing energy, energy consumption cannot be traded off by sacrificing age performance if the transmit power is constant. But if the transmit power can be adjusted, the tradeoff exists as power control provides a new degree of freedom. (4) The volume of sensing energy relative to transmit energy determines the range of the energy-age tradeoff.

This paper is a prior work on energy-age tradeoff. Future work can consider hybrid automatic repeat request (HARQ) strategy. Current policy simply abandons the failed packet and decodes a new one. If HARQ is applied, the failed packets can be reused and combined for decoding, so that the failure probability decreases with the increase of retransmission number. Thus, HARQ provides another degree of freedom to improve the energy-age tradeoff.



\bibliographystyle{IEEEtran}
\bibliography{ref}

\end{document}